\documentclass[twocolumn,british,prl,aps,showpacs,superscriptaddress]{revtex4-2}
\usepackage[letterpaper]{geometry}
\usepackage{url}
\usepackage{graphicx}
\usepackage{amsmath,bbm}
\usepackage{amssymb}
\usepackage{esint}
\usepackage{epstopdf}
\usepackage{color}
\usepackage{babel}
\usepackage{graphics}
\usepackage{epsfig}
\usepackage{bbold}

\newcommand{\ket}[2][]{{|#2\rangle_{#1}}}
\newcommand{\bra}[2][]{{}_{#1}\langle #2|}

\newcommand{\av}[1]{\langle #1\rangle}


\begin{document}

\title{Attaining classical capacity per unit cost of noisy bosonic Gaussian channels}
\author{Marcin Jarzyna}
\affiliation{Centre for Quantum Optical Technologies, Centre of New Technologies, University of Warsaw, ul. Banacha 2c, 02-097 Warszawa}
\date{\today}

\begin{abstract}
I show that classical capacity per unit cost of noisy bosonic Gaussian channels can be attained by employing generalized on-off keying modulation format and a projective measurement of individual output states. This means that neither complicated collective measurements nor phase-sensitive detection is required to communicate over optical channels at the ultimate limit imposed by laws of quantum mechanics in the limit of low average cost.
\end{abstract}

\maketitle

Transmission of information lies at the backbone of countless technologies. Utilizing quantum nature of light brings promise to increase information transmission rate beyond what is attainable by all conventional means \cite{Caves1994}. In particular, for a pure-loss channel in moderate and low power regimes, as measured by the average number of photons per time bin $n_a$, conventional receiver architectures like homodyne and heterodyne measurements, known to be almost optimal in the large power regime, are vastly outperformed by the ultimate quantum limit on the transmission rate, known as classical capacity \cite{Holevo1998, Schumacher1997}. The difference is of qualitative nature as conventional phase-sensitive detection schemes allow for rates scaling linearly with the average number of photons per time bin $\sim n_a$ in the small $n_a$ regime, whereas classical capacity scales as $\sim n_a\log_2\frac{1}{n_a}$ in the leading order \cite{Guha2011, Takeoka2014, Giovannetti2014}.

This discrepancy is even more evident when one looks at the capacity per unit 
cost (CPC) which quantifies the maximum amount of information that can be 
transmitted per single photon for a particular protocol, i.e. assuming certain modulation and detection scheme \cite{Verdu1990}. CPC contains all 
information about behavior of the conventional capacity in the regime of 
small power as the latter is just given by the CPC multiplied by the photon flux. 
Importantly, CPC indicates the maximum attainable value of the photon information 
efficiency (PIE) which is an important figure in many communication scenarios 
\cite{Stevens2008, Dolinar2011, Dolinar2012, Chandrasekaran2014, Takeoka2014, 
Zwolinski2018} and quantifies how much information can be transmitted per 
single photon for a given protocol operating at a particular level of $n_a$. In the case of a lossy channel conventional phase 
sensitive schemes allow for a constant CPC whereas the quantum-limited CPC is 
infinite \cite{Guha2011}. Although direct detection allows to attain the 
latter \cite{Jarzyna2015}, the second order term $\sim\log_2\log\frac{1}{n_a}$
 appearing in the respective PIE and lacking in the quantum-limitted PIE 
indicates a diverging difference between the two scenarios \cite{Chung2016}. 
For additive noise channels the maximum CPC allowed by laws of quantum 
mechanics is finite but still greater than  what can be attained by 
conventional receivers \cite{Jarzyna2017, Ding2019}.

It is known that in order to saturate the classical capacity or CPC of the pure-loss and most other physical channels, it is in principle necessary to use collective measurements on large number of channel outputs \cite{Hausladen1996, Sasaki1997, Bennett1997, Sasaki1998, Lloyd2011, Chung2016}, which are usually not feasible in practice. This is because of the superadditivity of accessible information with respect to joint measurements i.e. more information per channel use can be gained by considering collective measurements on several channel outputs than by detection of just a single output at a time \cite{Holevo1979, Peres1991}. Note that this is a different effect than superadditivity of the classical capacity \cite{Hastings2009} which is due to the possibility of using input states that are entangled between subsequent channel uses. Even coherent detection schemes operating on single symbols like the Kennedy \cite{Kennedy1973} or Dolinar \cite{Dolinar1973} receivers, despite outperforming classical shot noise limited receivers, cannot attain the capacity limit and have been realized only as proof of principle examples \cite{Cook2007, Becerra2013}. Another issue in reaching the optimal performance is that the required ensemble of input states is a continuous family of coherent states with a Gaussian prior distribution which may be problematic to produce in realistic applications. It is therefore crucial to identify measurement schemes and modulation protocols that attain the CPC bound and can be realized by current or near-future existing technology.

\begin{figure}[t]
\includegraphics[width=\columnwidth]{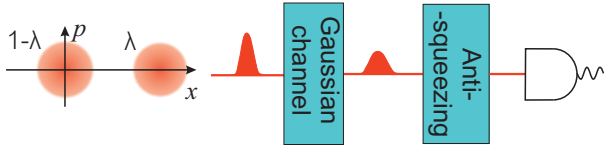}
\caption{A scheme presenting communication with generalized OOK modulation. A coherent state signal or a zero-cost state represented by an empty time bin are sent with probabilities $\lambda$ and $1-\lambda$ respectively. The states evolve through a Gaussian channel that changes the amplitude and adds noise. Detection stage implements a projective measurement onto eigenbasis of an output of a zero-cost state which can be realized by placing an antisqueezing operation  
and a subsequent photon number resolving measurement.}\label{Fig:OOKscheme}
\end{figure}

In this paper I show that a realistic single-symbol projective measurement is sufficient to asymptotically saturate CPC and  thus also classical capacity of any noisy Gaussian quantum channel in the low cost limit. The optimal signal modulation format is just a binary signal alphabet known as generalized on-off keying (OOK) and presented schematically in Fig.~\ref{Fig:OOKscheme}. It is composed from an empty time bin (vacuum state) and an infrequently sent signal in a coherent state. Up to my knowledge this is the first result in which it is shown how to attain classical capacity of realistic communication channels without using highly complicated receiver architecture employing sophisticated quantum measurements. Note that some partial results on the attainability of CPC were obtained in \cite{Jarzyna2017} and \cite{Ding2019} in which it was shown respectively that binary or pulse position modulation formats are enough, however, the optimal POVMs were found either only in the trivial case of the noiseless channel or were collective.

For a communication setup described by a quantum channel $\Lambda$ with a given ensemble of input states $\rho_x$ used with an input symbol probability distribution $p(x)$ and receiver performing a measurement described by a positive operator valued measure (POVM) $\{\Pi_y\}$ the maximal information transmission rate is quantified by the mutual information $I=H(Y)-H(Y|X)$, where $H(Y)=-\sum_y p(y)\log_2 p(y)$ and $H(Y|X)=-\sum_x\sum_y p(x)p(y|x)\log_2 p(y|x)$ with conditional probability distribution evaluated through the Born rule $p(y|x)=\textrm{Tr}(\Lambda[\rho_x] \Pi_y)$. Importantly, each use of one of the states $\rho_x$ is usually assumed posses some kind of cost, determined by a properly chosen cost function $b[\rho_x]$, e.g. the energy of the state. Mutual information optimized over input symbol probability distributions is known as the capacity of the channel and quantifies the best rate for a given communication protocol, i.e. assuming particular modulation and detection schemes. Capacity further optimized over input states ensembles and, possibly collective, measurements performed on many channel outputs returns the maximal information transmission rate for the channel, known as the classical capacity $C$. Finding classical capacity is in general a formidable task, however, for a special class of channels that I will consider, known as Gaussian channels, this problem can be solved under the constraint of fixed average cost per channel use $n_a=\sum_x p(x)b[\rho_x]$ \cite{Giovannetti2004a, Giovannetti2014, Schafer2016}.

A convenient quantity to analyze communication in the low cost regime is CPC which is given by the maximum ratio of the capacity and average cost per channel use. CPC characterizes the maximum amount of information that can be carried out per unit cost i.e. the efficiency of communication rather than just the information rate. In the case of Gaussian channels the usual cost figure is the average energy of the state and I will focus on this scenario. For a quantum channel that allows to use a a zero-cost state, or in other words a vacuum  $\rho_0=\ket{0}\bra{0}$, the CPC can be expressed by a compact formula \cite{Verdu1990}
\begin{equation}\label{eq:cpc_class}
\mathbf{C}_{\textrm{class}}=\max_{x\neq 0}\frac{D\left[p(y|x)||p(y|0)\right]}{n_s},
\end{equation} 
where $n_s=b[\rho_x]$ is the cost of state $\rho_x$ and $D[p(y|x)||p(y|0)]=\sum_y p(y|x)\log_2\frac{p(y|x)}{p(y|0)}$ is the Kullback-Liebler divergence between distributions $p(y|x)$ and $p(y|0)=\textrm{Tr}(\Lambda[\rho_0]\Pi_y)$. Note that maximization in eq.~(\ref{eq:cpc_class}) is taken over input symbols  but not the input states since the set of the latter is considered fixed. If one adds optimization over input state ensembles and measurements on top of eq.~(\ref{eq:cpc_class}) one obtains the classical capacity per unit cost (CCPC) $\mathbf{C}_{\textrm{quant}}$ which quantifies the best possible PIE attainable for a given quantum channel \cite{Jarzyna2017, Ding2019} and is equal to
\begin{equation}\label{eq:cpc_quant_gen}
\mathbf{C}_{\textrm{quant}}=\max_{\rho\neq \ket{0}\bra{0}}\frac{D\left[\Lambda[\rho]||\Lambda[\rho_0]\right]}{n_s}.
\end{equation}
Both CPC and CCPC are attained in the limit of vanishing average cost per channel use $n_a\to 0$ \cite{Verdu1990, Jarzyna2017, Ding2019}. Therefore, since they are defined as maximum ratios of respective capacities per channel use and average cost, in the low cost limit one obtains for the classical capacity $C\approx n_a\mathbf{C}_{\textrm{quant}}$ and similarly for the regular capacity. Importantly, in the low cost regime, CCPC is exactly equal to the PIE of the optimal protocol saturating the classical capacity of the channel.

A basic model of an optical communication channel is a Gaussian bosonic channel. Gaussian channels describe various effects that are characterized by evolution quadratic in creation and annihilation operators of the system, such as linear losses, thermal noise, phase sensitive noise or squeezing \cite{Weedbrook2012}. They can be characterized as the most general type of operations that preserve Gaussian character of quantum states on which they act. 
A general Gaussian channel can be specified by a real matrix $\bold{X}$ and a real symmetric and nonnegative matrix $\bold{Y}$ which satisfy certain conditions \cite{Holevo2001} to ensure complete positivity and trace preservation by the channel. The output state first moments and covariance matrix are given by
\begin{equation}
\bold{d}_{\textrm{out}}=\bold{X}\bold{d}_{\textrm{in}},\quad \bold{V}_{\textrm{out}}=\bold{X}\bold{V}_{\textrm{in}}\bold{X}^T+\bold{Y},
\end{equation}
where $\bold{d}_\textrm{in}$ and $\bold{V}_\textrm{in}$ are respectively the first moments vector and the covariance matrix of the input state. Any Gaussian channel can be decomposed into a fiducial channel and passive (i.e. conserving the energy) Gaussian unitary operations preceding and following the former \cite{Schafer2013}. For a channel specified by $\bold{X}$ and $\bold{Y}$ the fiducial channel is given by matrices
\begin{equation}\label{eq:fiducial}
\bold{X}_F=\sqrt{|\eta|}\left(
\begin{array}{cc}
1 & 0\\
0 & \textrm{sgn}(\eta)
\end{array}\right),\quad 
\bold{Y}_F=y\left(
\begin{array}{cc}
e^{2s} & 0 \\
0 & e^{-2s}
\end{array}
\right),
\end{equation}
where $\eta=\det\bold{X}$, $y=\sqrt{\det\bold{Y}}$ and $s$ can be interpreted as intrinsic channel squeezing. The original channel matrices can be written as
\begin{equation}\label{eq:fiducial_m}
\bold{X}=\bold{M}\bold{X}_F\bold{\Theta},\quad \bold{Y}=\bold{M}\bold{Y}_F\bold{M}^T,
\end{equation}
where $\bold{M}$ is some symplectic operation and $\bold{\Theta}$ denotes phase space rotation. The exact relations between matrices in this decomposition and the original channel can be found in \cite{Schafer2013}. Parameter $\eta$ can be interpreted as a characteristic transmission coefficient of the channel, note, however, that it can be a number with an absolute value larger than $1$ which describes phase-conjugating and amplifying channels. 

The CCPC of any Gaussian channel is equal to CCPC of its corresponding fiducial channel. This is because the phase-space rotation $\bold{\Theta}$ in Eq.~(\ref{eq:fiducial_m}) does not change energy of input states and the symplectic transformation $\bold{M}$ can be always undone by incorporating a proper unitary transformation at the channel output. 
It is known that for any Gaussian channel CCPC is saturable in the low cost limit by the generalized OOK modulation format \cite{Jarzyna2017}. For general quantum Gaussian channels it was 
shown in \cite{Jarzyna2017, Ding2019} that CCPC is equal to 
\begin{equation}\label{eq:cpc_gauss}
\mathbf{C}_{\textrm{quant}}=|\eta|\omega_{\max}\log_2\left(1+\frac{1}{n_b}\right),
\end{equation}
where the parameter $\eta$ is defined in Eq.~(\ref{eq:fiducial}) and $n_b$ and $\omega_{\max}=e^{2r}$ are respectively the average thermal energy and the squeezing of the output state of the fiducial channel if the input was in a vacuum state
\begin{gather}
n_b+\frac{1}{2}=\sqrt{\left(\frac{|\eta|}{2}+ye^{2s}\right)\left(\frac{|\eta|}{2}+ye^{-2s}\right)},\\
\omega_{\max}=e^{2r}=\sqrt{\frac{\frac{|\eta|}{2}+ye^{2s}}{\frac{|\eta|}{2}+ye^{-2s}}}.
\end{gather}
Importantly, for any Gaussian channel 
with additive noise CCPC has a finite value, meaning that in the low cost regime 
classical capacity is equal to $C\approx n_a \mathbf{C}$. Therefore, it is enough to show 
that a receiver attains the CCPC to show its optimality also from the point 
of view of the actual capacity per channel use.

I will consider the generalized OOK modulation format, shown schematically in Fig.~\ref{Fig:OOKscheme}. The input message is encoded in a series of time bins, each of which can be either empty with probability $1-\lambda$ or can carry a coherent state with an average energy $n_s$ with corresponding probability $\lambda$. The average cost per channel use of such an input ensemble is given by $n_a=\lambda n_s$. The phase space rotation $\bold{\Theta}$ appearing in the decomposition Eq.~(\ref{eq:fiducial_m}) changes just the phase of the input state and thus can be neglected without loss of generality by tuning the phase of the coherent state properly. For input in a coherent state with amplitude $\alpha$ the output of the fiducial channel of a general single-mode Gaussian channel is given by a density matrix
\begin{equation}\label{eq:coherent}
\rho=\hat{D}\left(\sqrt{|\eta|}\alpha\right)\hat{S}(r)\rho_{n_b}\hat{S}^{\dagger}(r)\hat{D}^{\dagger}\left(\sqrt{|\eta|}\alpha\right),
\end{equation}
where $\hat{D}(.),\,\hat{S}(.)$ are displacement and squeezing operators and $\rho_{n_b}=\sum_{k=0}^\infty\frac{n_b^k}{(n_b+1)^{k+1}}\ket{k}\bra{k}$ is a thermal state with the average energy $n_b$ and I have chosen the phase of $\alpha$ such that the state is aligned with the position quadrature.

Order of the squeezing and displacement operators in eq.~(\ref{eq:coherent}) can be exchanged by the rule \cite{Marian1993}
\begin{equation}\label{eq:gamma}
\hat{D}(\sqrt{|\eta|}\alpha)\hat{S}(r)=\hat{S}(r)\hat{D}(\gamma),\quad \gamma=\sqrt{|\eta|}\alpha e^{r}.
\end{equation}
Note, that since $r$ for a fiducial channel is defined to be a real parameter, the phase of $\gamma$ depends only on the phase of $\alpha$, which I set to zero.
Therefore the  output state in Eq.~(\ref{eq:coherent}) can be written as
\begin{equation}\label{eq:rho_exhange}
\rho=\hat{S}(r)\hat{D}(\gamma)\rho_{y}\hat{D}^{\dagger}(\gamma)\hat{S}^{\dagger}(r).
\end{equation}
The state in Eq.~(\ref{eq:rho_exhange}) undergoes then evolution through the unitary transformation  $\hat{U}_\bold{M}$ corresponding to symplectic operation $\bold{M}$ in Eq.~(\ref{eq:fiducial_m}) which gives eventually
\begin{equation}\label{eq:rho_final}
\rho=\hat{U}_{\bold{M}}\hat{S}(r)\hat{D}(\gamma)\rho_{y}\hat{D}^{\dagger}(\gamma)\hat{S}^{\dagger}(r)\hat{U}_{\bold{M}}^\dagger.
\end{equation}
Importantly, any one-mode symplectic transformation can be realized by a combination of two phase space rotations and squeezing. Therefore, one can write $\hat{U}_{\bold{M}}=\hat{U}_{\theta_2}\hat{S}(z)\hat{U}_{\theta_1}$, where $\hat{U}_{\theta_i}$ denotes the unitary rotation by phase $\theta_i$ and $z$ is the squeezing introduced by $\bold{M}$.

The POVM that I will consider is a projective measurement onto the eigenbasis of an output of the zero-cost state. In case of Gaussian channels with average number of photons per channel use constraint the latter is  the vacuum state's output, given by eq.~(\ref{eq:rho_final}) with $\gamma=0$. The measurement is therefore given by projections onto squeezed number states $\Pi_k=\hat{U}_{\theta_2}\hat{S}(z)\hat{U}_{\theta_1}\hat{S}(r)\ket{k}\bra{k}\hat{S}^{\dagger}(r)\hat{U}_{\theta_1}^\dagger\hat{S}^\dagger(z)\hat{U}_{\theta_2}^\dagger$. It can be experimentally realized by a photon number resolving measurement preceded by an antisqueezing operation in the right direction or by a proper combination of phase space rotations and squeezing, see Fig.~(\ref{Fig:OOKscheme}). 

For the coherent state with amplitude $\alpha$ the measurement statistics is given  by a conditional probability distribution \cite{Marian1993}
\begin{equation}\label{eq:probability}
p(k|\alpha)=\frac{n_b^k}{(n_b+1)^{k+1}}e^{-\frac{|\gamma|^2}{n_b+1}}L_k\left(-\frac{|\gamma|^2}{n_b(n_b+1)}\right),
\end{equation}
where $L_k$ denotes the $k$th Laguerre polynomial and $\gamma$ is given by eq.~(\ref{eq:gamma}). In the generalized OOK modulation format $\alpha$ can take only two values: $\alpha_0=0$ for the vacuum state and $\alpha_s=\sqrt{n_s}$ for the signal state. Plugging eq.~(\ref{eq:probability}) into eq.~(\ref{eq:cpc_class}), the capacity per unit cost is equal to
\begin{multline}\label{eq:cpc_pnr}
\mathbf{C}_{\textrm{class}}=\frac{1}{n_s}\left\{\sum_{k=0}^\infty p(k|\sqrt{n_s})\log_2 \frac{1}{p(k|0)}+\right.\\-H[p(k|\sqrt{n_s})]\Bigg\},
\end{multline}
where $H[p(x)]=-\sum_x p(x)\log_2 p(x)$ is the Shannon entropy of distribution $p(x)$. The first term in the bracket in eq.(\ref{eq:cpc_pnr}) is equal to
\begin{multline}\label{eq:first_term}
\sum_{k=0}^\infty p(k|\sqrt{n_s})\log_2 \frac{1}{p(k|0)}=\\(n_b+|\gamma_s|^2)\log_2\left(1+\frac{1}{n_b}\right)+\log_2(1+n_b),
\end{multline}
where $\gamma_s$ is the displacement defined in eq.~(\ref{eq:gamma}) evaluated for $\alpha_s=\sqrt{n_s}$. Since $|\gamma_s|^2=|\eta|n_s\omega_{\max}$ in eq.~(\ref{eq:gamma}), for large $n_s$ the expression in eq.~(\ref{eq:first_term}) is  in the leading order equal to $|\eta| n_s\omega_{\max} \log_2\left(1+\frac{1}{n_b}\right)$, which is exactly the CCPC of the channel multiplied by cost of the signal state. The remaining term in the bracket in eq.~(\ref{eq:cpc_pnr}) is upper bounded by
\begin{equation}\label{eq:ent_pnr}
H[p(k|\sqrt{n_s})]\leq g(n_b+|\gamma_s|^2),
\end{equation}
where $g(x)=(x+1)\log_2(x+1)-x\log_2(x)$ is the entropy of a thermal state with an average energy $x$. This is because the right hand side is the maximal possible entropy for any distribution with a fixed expected value $\av{k}=n_b+|\gamma_s|^2$. Since $|\gamma_s|^2\sim n_s$ in eq.~(\ref{eq:gamma}) in large $n_s$ limit the expression on the right hand side of eq.~(\ref{eq:ent_pnr}) is equal to $\log_2|\gamma_s|^2+O(1)$. Plugging these results into eq.~(\ref{eq:cpc_pnr}) and going with signal cost to infinity $n_s\to\infty$ one obtains that the term corresponding to eq.~(\ref{eq:ent_pnr}) vanishes and the capacity per unit cost is equal to the quantum bound in eq.~(\ref{eq:cpc_gauss}), i.e.
\begin{equation}\label{eq:class_quant}
\mathbf{C}_{\textrm{class}}=\mathbf{C}_{\textrm{quant}}=|\eta|\omega_{\max}\log_2\left(1+\frac{1}{n_b}\right).
\end{equation}

\begin{figure}[t]
\includegraphics[width=\columnwidth]{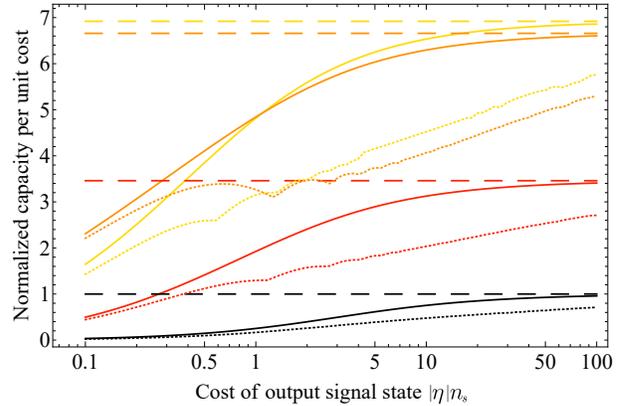}
\caption{(Color online) Capacity per unit cost normalized to channel transmission $|\eta|$ as a function of the cost of the output signal state $|\eta| n_s$ for generalized OOK modulation and projective measurement onto appropriately squeezed number state (solid curves), threshold detector (dotted curves) and the ultimate quantum bound given by classical capacity per unit cost (dashed lines). Results for phase-insensitive channels are $n_b=1$ - black; $n_b=0.1$ - red (dark grey), $n_b=0.01$ - orange (gray) while the phase sensitive channel with $n_b=0.1$ and $r=\ln2 /2\approx 0.34$ is depicted by yellow (light gray) curves.}\label{Fig:pie_na}
\end{figure}

Note that the result in eq.~(\ref{eq:class_quant}) is asymptotic, i.e. it applies in the limit of vanishing average cost per time bin $n_a\to 0$, since this is the limit in which capacities per unit cost can be attained. However, if one choses an average cost per time bin at the output $|\eta|n_a$ as a figure instead, the conclusions remain unchanged, i.e. the capacity per unit cost is still equal to eq.~(\ref{eq:class_quant}) but normalized to the transmission $|\eta|$ and it is saturated for small output average cost per time bin $|\eta|n_a\to0$. The CPC in Eq.~(\ref{eq:class_quant}) diverges logarithmically with decreasing additive noise $n_b$ meaning that for channels that do not introduce any additive noise, such as pure-loss and lossless channels, there is no limit on the attainable CPC. This was observed in \cite{Guha2011,  Chen2012, Jarzyna2015, Chung2016} where it was shown that direct detection together with generalized OOK in principle allows attaining unbounded CPC for lossy channels.

Fig.~\ref{Fig:pie_na} shows classical capacity per unit cost normalized to the channel transmission $|\eta|$ obtained for the discussed communication strategy as a function of the cost of the output signal state. It is seen that respective capacities converge to the asymptotic values given by the ultimate quantum limit Eq.~(\ref{eq:cpc_gauss}) both for phase-sensitive and phase-insensitive channels. In the considered range of phase-insensitive noise $n_b=0.01-1$ the saturation happens around the signal output cost  $|\eta|n_s\approx 10$ which confirms the assumption of a necessary large signal state cost. The latter is always possible since $n_a=\lambda n_s$ and the probability of sending coherent state can be freely chosen. It is seen in Fig.~(\ref{Fig:pie_na}) that for full saturation of the CCPC it is necessary to send strong coherent states $n_s\to\infty$ very rarely $\lambda\to0 $. Note, that this means that the total transmission time of a message may be large, although this is expected since in small $n_a$ regime the capacity is low. From a practical point of view, this time may be decreased by increasing the bandwidth of the link.

The detection scheme I proposed above assumed no dependence on the average cost. IF such dependence is allowed one may propose an even simpler binary detection scheme that still allows to attain CCPC asymptotically. Consider a threshold detector with two-component POVM $\Pi_0^{(\textrm{th})}=\sum_{k=0}^{k^{(\textrm{th})}}\Pi_k$ and $\Pi_1^{(\textrm{th})}=\mathbb{1}-\Pi_0^{(\textrm{th})}$. The threshold value $k^{(\textrm{th})}$ can be freely adjusted by the receiver to the average cost of the incoming signal. With such choice, one obtains two outcomes with respective probabilities
\begin{equation}\label{eq:po}
p_0(\alpha)=\sum_{k=0}^{k^{(\textrm{th})}}p(k|\alpha),\quad p_1(\alpha)=1-p_0(\alpha),
\end{equation}
where $p(k|\alpha)$ are defined in Eq.~(\ref{eq:probability}). Plugging the above distribution evaluated for the signal state and the zero-cost state into Eq.~(\ref{eq:cpc_class}) one gets
\begin{multline}\label{eq:capth_1}
\mathbf{C}^{(\textrm{th})}=\frac{1}{n_s}\Bigg\{h_2[p_0(\sqrt{n_s})]+\\+\left[p_0(\sqrt{n_s})\log_2\frac{1}{p_0(0)}+p_1(\sqrt{n_s})\log_2\frac{1}{p_1(0)}\right]\Bigg\},
\end{multline}
where $h_2(x)$ denotes the binary entropy function. The first term in Eq.~(\ref{eq:capth_1}) has to be smaller than the entropy on the left hand side of Eq.~(\ref{eq:ent_pnr}) due to the data processing inequality and thus gives a vanishing contribution to CPC for large $n_s$. The distribution in Eq.~(\ref{eq:probability}) converges to a Gaussian distribution in the large $n_s$ limit and its variance is equal to $\textrm{Var}(k)=|\gamma_s|^2(1+2n_b)+n_b(n_b+1)$. Therefore if one takes $k^{(\textrm{th})}=(1-\epsilon)|\gamma_s|^2$ with any $\epsilon>0$ one can always find a coherent state cost value $n_1$ such that for any $n_s>n_1$ one has $k^{(\textrm{th})}< \mathbb{E}(k)-c\sqrt{\textrm{Var}(k)}$, where $\mathbb{E}(k)=|\gamma_s|^2+n_b$ is the expectation value of the distribution in Eq.~(\ref{eq:probability}) and $c$ is an arbitrary constant. This means that for sufficiently large $n_s$ the probability $p(k|\sqrt{n_s})$ takes non-negligible values only for $k>k^{(\textrm{th})}$. As a consequence, $p_1(\sqrt{n_s})\approx 1$ and $p_0(\sqrt{n_s})\approx 0$, which gives
\begin{multline}\label{eq:treshold}
\mathbf{C}^{(\textrm{th})}\approx \frac{k^{(\textrm{th})}+1}{n_s}\log_2\left(1+\frac{1}{n_b}\right)=\\=\frac{(1-\epsilon)|\gamma_s|^2+1}{n_s}\log_2\left(1+\frac{1}{n_b}\right),
\end{multline}
where I have plugged the expression for $p_1(0)$ from eq.~(\ref{eq:po}). Since Eq.~(\ref{eq:treshold}) is valid for any $\epsilon>0$, CPC has to converge to the ultimate quantum limit Eq.~(\ref{eq:cpc_quant_gen}) with $n_s\to\infty$. It is seen in Fig.~(\ref{Fig:pie_na}) that threshold detector also allows for saturation of the quantum CPC bound, although necessary signal cost $n_s$ is much larger than for projection onto zero-cost state output eigenvectors. Unlike the previous case, the CPC for the threshold detector is not a monotonic function of $n_s$ because of the discrete nature of the threshold $k^{(\textrm{th})}$.

In conclusion I have showed that the ultimate quantum bound on the information transmission rate per unit of energy of a general noisy bosonic Gaussian channel can be asymptotically attained by the generalized OOK modulation and a projective measurement on the individual channel outputs in the form of projection onto squeezed number states. As an implication the classical capacity of noisy Gaussian channels in the low cost limit can also be saturated by the considered protocol. This is a qualitatively different situation than in the pure-loss case in which there appears a non-vanishing gap between the classical channel capacity and what can be achieved with individual measurements.

I thank K. Banaszek, L. Kunz, W. Zwoli\'{n}ski and M. Lipka for insightful discussions. This work was supported by the Foundation for Polish Science under the TEAM project ``Quantum Optical Communication Systems'' co-financed by the European Union under the European Regional Development Fund.

\end{document}